\documentclass[aps,pra,groupedaddress,floatfix,twocolumn,showpacs]{revtex4}
\usepackage{graphicx}
\usepackage{dcolumn}
\usepackage{bm}
\usepackage{amsmath}
\usepackage{amsfonts}
\usepackage{amssymb}
\usepackage{longtable}

\begin{document}
\title{Scaling the neutral atom Rydberg gate quantum computer by collective encoding in Holmium atoms}
\author{M. Saffman$^1$ and K. M\o lmer$^2$}

\affiliation{$^{1}$Department of Physics, University of Wisconsin,
1150 University Avenue, Madison, Wisconsin 53706, USA}
\date{\today}

\affiliation{$^{2}$Lundbeck Foundation Theoretical Center for
Quantum System Research, Department of Physics and Astronomy,
University of Aarhus, DK-8000 \AA rhus C, Denmark}

\begin{abstract}
We discuss a method  for scaling a neutral atom Rydberg gate quantum processor to a large number of qubits. 
Limits are derived showing that  the number of qubits that can be directly connected by entangling gates with errors at the $10^{-3}$ level
using long range Rydberg interactions between sites in an optical lattice, without mechanical motion or swap chains, is about 500 in two dimensions and 7500 in three dimensions. A  scaling factor of  60 at a smaller number of sites can be obtained using collective 
register encoding in the hyperfine ground states of the rare earth atom Holmium. We present a detailed analysis of operation of the 60 qubit register in Holmium. Combining 
a lattice of multi-qubit ensembles with collective encoding results in a feasible design for a 1000 qubit fully connected quantum processor. 
\end{abstract}
\maketitle

\section{Introduction}

Quantum computing has the potential for performing numerical calculations such as 
factoring and unstructured search faster than is 
possible on classical computers\cite{nielsenchuang}. Despite
the slower speed of basic gate operations on quantum compared to classical computers
an algorithmic speedup is predicted for sufficiently large problems. 
Impressive yet rudimentary demonstrations of small quantum algorithms using less than 10 qubits 
have been achieved using several 
different physical embodiments of quantum bits and gates. These include controlled entanglement of up to 8 
trapped ions\cite{ref.wineland,ref.blatt}, and factoring with molecular spins manipulated by 
nuclear magnetic resonance techniques\cite{ref.chuang}. In order to realize the full potential of quantum algorithms for
solving difficult computational problems it will be necessary to develop approaches which 
allow a large number of qubits to be interconnected. Several conceptual designs for scalable 
quantum computing architectures have appeared in recent years based on ion traps\cite{ref.scalableion}, and there is intense effort directed at scalability of other approaches including quantum 
dots\cite{ref.scalableqd}, superconductors\cite{ref.scalablesc}, linear optics\cite{ref.scalablelo}, rare earth crystals\cite{ref.scalablerareearth}, and small quantum repeaters\cite{ref.scalablerepeater}.

In this paper we examine the potential scalability of a computer based on neutral atom qubits trapped in optical lattices with long range two-qubit gates mediated by dipolar interactions of Rydberg states\cite{ref.jaksch}. We will carefully estimate the maximum number of qubits that can be directly interconnected. 
Although it is in   principle possible to move information arbitrary distances in a quantum computer by swap operations between neighboring qubits, there is strong evidence for an increase in the error threshold compatible with scalable computations when swap operations are required\cite{ref.divincenzo}. Conversely it is possible to put rigorous bounds on the required error threshold  in globally connected 
models\cite{ref.preskill}. An alternative approach to coupling  qubits separated by distances that exceed the range of a direct interaction is based on conversion of the quantum information from stationary to flying and then back to stationary qubits. This potentially eliminates the need for a long chain of swap operations but introduces the nontrivial requirement of a high efficiency stationary - flying qubit interface for most of the qubits in the quantum processor. 

While we do not dispute the potential of the above approaches to long range interconnections it is interesting to explore how large a system one might build that allows each pair of qubits to be directly entangled without intermediate steps. 
We present here an analysis that takes advantage of the long range of Rydberg atom dipole-dipole interactions
as well as the scaling factor provided by collective addressing\cite{ref.molmer1,ref.molmer2}. 
We divide our discussion into several parts. In Sec. \ref{sec.lattice} we make estimates of the number of 
qubits that can be directly connected in an optical lattice using Rydberg interactions. 
In Sec. \ref{sec.holmium} we show that collective encoding in the rare earth atom Ho can potentially provide  a scaling factor as large as 60 per site. In Sec. \ref{sec.collisions} we discuss a protocol for filling an array of localized ensembles in a planar lattice, and  estimate the total size of the globally connected processor. We conclude in Sec. \ref{sec.discussion} with a summary and outlook for the future.

\section{Rydberg gates in optical lattices}
\label{sec.lattice}

The idea of using the dipole-dipole interaction of Rydberg atoms for neutral atom quantum gates was introduced 
by Jaksch and coworkers in \cite{ref.jaksch} and subsequently extended to a mesoscopic encoding of the 
qubit by Lukin and coworkers\cite{ref.lukin}. A number of subsequent papers have analyzed in more detail specific schemes for implementing  Rydberg gates\cite{ref.grangier,ref.swth1,ref.zollerfast} with the conclusion that they present a very promising approach to quantum logic. At this time a neutral atom Rydberg gate has not been demonstrated experimentally, although many of the underlying requirements have been separately achieved including loading and manipulation of single atoms in optical traps\cite{ref.meschede,ref.grangiersa,ref.wiscexp1}, signatures of Rydberg interactions and dipole blockade at the many atom 
level\cite{ref.dipoleblockade}, and recently coherent excitation of Rydberg states together with observation of interaction effects at the level 
of two atoms\cite{ref.wiscexp2}.  

\begin{figure}
\centering
\includegraphics[width=7cm]{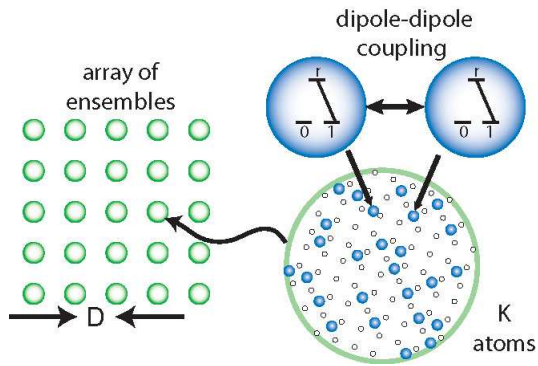}
\caption{(color online) Planar optical lattice defining an array of sites with spacing  $D$, each one of which may be a
single atom or a  small ensemble containing $K$ atoms. Rydberg coupling enables gates within each ensemble and betweeen ensembles.}
\label{fig.array}
\end{figure}

We envision an array of atoms in optical traps defining the spatial geometry of the quantum computer as shown in Fig. \ref{fig.array}. 
The number of qubits that can be directly connected depends on the dimensionality of the confining lattice. 
A three dimensional lattice offers the benefit of a higher packing density but  it is also  substantially more difficult to manipulate and measure the qubit state at a single site without disturbing neighboring sites in a three dimensional geometry. We will therefore concentrate on a two-dimensional array of sites that can be individually addressed. 
Quite arbitrary arrays of trapping sites can be readily produced with diffractive optical elements or spatial light modulators using known techniques. We initially assume that  each site contains a single atomic qubit and then extend this to allow for a three dimensional sublattice within each site which is collectively addressed and contains an ensemble of $K$ atoms that encode an 
$N$ qubit register.  

\subsection{Maximum number of connected qubits in a lattice}

A two-qubit gate between sites 1 and 2 separated by a distance $R$ is achieved by exciting atoms in the two sites to a high lying Rydberg level with principal quantum number $n.$ 
When the sites are separated by a relatively large distance $R$  we have a van der Waals interaction of the form $U_{\rm vdW}=C_6/R^6$. The asymptotic scaling at large $n$ is\cite{ref.gallagherbook} 
\begin{equation}
C_6\simeq \frac{1}{(4\pi\epsilon_0)^2}\left(\frac{3}{2}\right)^4 \frac{q^4 a_0^4}{ E_{\rm R}} n^{11}.
\label{eq.C6}
\end{equation} 
Here $q$ is the electronic charge, $a_0$ is the Bohr radius, and $E_{\rm R}$ is the Rydberg energy. 
The $n^{11}$ scaling seen in Eq. (\ref{eq.C6}) should be valid for all atomic species. However, this estimate turns out to be  too conservative since it is based on assuming the F\"orster energy defect scales  
as $\delta\sim E_{\rm R}/n^3$. In the heavy alkalis the difference between the $s$ and $p$ quantum defects is close to $0.5$ which results in near cancelation of the $1/n^3$ dependence. We find that numerical calculations of $\delta$ and $C_6$\cite{ref.ws2008} are reproduced much better for $50<n<200$ by 
using $\delta= 5E_{\rm R}/n^4$ which implies 
 $C_6\sim n^{12}$. This scaling will be obtained for Rydberg $s$ states whenever the difference of the $s$ and $p$ quantum defects is an integer, half integer, or quarter integer. In addition, even when the quantum defects do not conspire to minimize the F\"orster energy, external fields can be used to Stark and/or Zeeman tune the levels into resonance. In light of these considerations we will assume that $\delta=k_\delta E_{\rm R}/n^4$, with $k_\delta$ a scaling constant, 
which gives  
\begin{equation}
C_6\simeq \frac{1}{(4\pi\epsilon_0)^2}\frac{1}{k_\delta}\left(\frac{3}{2}\right)^4 \frac{q^4 a_0^4}{ E_{\rm R}} n^{12}.
\label{eq.C6p}
\end{equation} 
Although we will be interested in what follows in Ho atoms, due to lack of detailed knowledge of the Ho Rydberg spectrum and excited state lifetimes, we base the numerical estimates  given in this section on Rb. 

To find the maximum number of qubits that can be directly connected using Rydberg interactions we first estimate the maximum separation $R_{\rm max}$ for a desired gate error $E.$ In the van der Waals limit we assume the excitation Rabi frequency $\Omega$ is large compared to the interaction frequency $\Delta_{\rm vdW}=U_{\rm vdW}/\hbar.$ In this limit, neglecting small corrections due to the finite energy separation of the hyperfine ground states, the minimum gate error averaged over all two-qubit input states is\cite{ref.swth1}
$E=\frac{3\pi^{2/3}}{2^{1/3}}\frac{1}{( \Omega \tau)^{2/3}},$ with $\tau$ the Rydberg lifetime.   
At room temperature the blackbody background limits the Rydberg lifetime to   $\tau\simeq \tau_0 n^2$ so the gate error scales as $E\sim n^{-4/3}.$ 
This error is achieved when the interaction strength is set to the optimum value $\Delta_{\rm opt}=\frac{}{}\left(\frac{\pi}{4}\right)^{1/3}\frac{\Omega^{2/3}}{\tau^{1/3}}=\frac{3\pi}{8^{1/3}}\frac{1}{\tau E}.$ Using the van der Waals scaling the maximum qubit separation at a fixed error is thus
\begin{equation}
R_{\rm max}=\left(\frac{2}{3\pi} \right)^{1/6} \left( \frac{C_6\tau E}{\hbar}\right)^{1/6}\sim n^{7/3}.
\end{equation}

Given $R_{\rm max}$ the number of sites that can be connected scales as $(R_{\rm max}/D_{\rm min})^d$ in $d$ dimensions with $D_{\rm min}$ the minimum usable lattice spacing. The minimum spacing is determined by several limits. The sites must be optically resolvable in order to perform gate operations on a desired qubit without disturbing neighboring qubits. 
In a lattice geometry there is an additional limitation related to the fact that we must avoid interactions between the highly excited electron of a  Rydberg atom  and a ground state atom at a neighboring site. This requirement can be written as $D_{\rm min}= k_1  a_0 n_{\rm max}^2$ with a safety factor $k_1 > 1.$ 
For alkali atom s states the radial wavefunction scales as 
$$
 R_{n^*,0}\sim e^{-r/a_0 n^*} U(1-n^*,2,2r/a_0 n^*).
$$
with $n^*$ the effective principal quantum number and $U$ the confluent hypergeometric function.  
At  large $r$,  $R_{n^*,0}\sim e^{-r/a_0 n^*} (r/a_0 n^*)^{n^*-1}$
 which has a maximum at $r_{n^*}=a_0(n^*)^2.$ At a larger distance $r'=k_1 r_{n^*}$ the wavefunction is less than its maximum value  by a factor of $e^{n^*(k_1-1)}/k_1^{n^*-1}.$ Setting this factor to $10^2$ (so the probability density is reduced by a factor $10^{4}$) and $n^*=100$ we find $k_1=1.32$ or 
$D_{\rm min}=0.7 ~\mu\rm m$.  Allowing for an additional positional uncertainty in the plane of the lattice for each atom 
of $0.15~\mu\rm m$ suggests a  lower limit of  $D_{\rm min}\simeq 1.0~\mu\rm m$, which corresponds 
to $k_{1,\rm eff}=1.89$.  
This value of $D_{\rm min}$ is consistent with the single site addressability requirement
using visible and near infrared lasers for internal state manipulation, provided fast diffraction limited addressing optics and ``top-hat" shaped beams are used.  

With these considerations in mind we find the maximum number of interconnected sites when relying on a $1/R^6$ van der Waals interaction in  square or cubic lattices is 
\begin{subequations}
\begin{eqnarray}
N_{\rm max,vdW}^{(\rm 2D)} &=&  \frac{\frac{\pi}{4}R_{\rm max}^2}{D_{\rm min}^2}\nonumber\\
&=&\frac{3\pi^{2/3} }{2^{8/3}k_1^2 k_\delta^{1/3} }\left(\frac{\alpha^2 m c^2 \tau_0}{ \hbar} \right)^{1/3}
\hspace{-.3cm}E^{1/3}n^{2/3}
\\
N_{\rm max,vdW}^{(\rm 3D)} &=& \frac{\frac{\pi}{6}R_{\rm max}^3}{D_{\rm min}^3}\nonumber\\
&=& \frac{3^{1/2}\pi^{1/2} }{4 k_1^3 k_\delta^{1/2}}\left(\frac{\alpha^2 m c^2 \tau_0}{ \hbar} \right)^{1/2}
E^{1/2} n.
\end{eqnarray}
\label{eq.Nmax2}
\end{subequations}
We see that the number of connected qubits at a fixed value of the gate error scales  $\sim n^{2/3}$ in 2D and $\sim n$ in 3D.  Putting in $k_1=1.89$, $k_\delta=5$, $\tau_0=54~\rm ns$, a target error of $E=0.001$, and $n=100$ gives 
$N_{\rm max,~vdW}^{(\rm 2D,3D)}=470,\, 7600$ with an array  
that is about $22~\mu\rm m$ on a side. 
These numbers can be increased further by increasing $n$, but there are some practical limitations to letting $n$ be arbitrarily large. These include  the difficulty of rapid laser excitation of high lying Rydberg states as well as sensitivity of the Rydberg states 
to external fields. Excitation of specific levels with $n$ as large as 500 has been achieved with careful shielding\cite{ref.dunning}. We will assume a conservative limit of $n_{\rm max}=100$ which ensures sufficiently long Rydberg radiative lifetimes that gate errors  ${\mathcal O}(10^{-3})$ are feasible\cite{ref.swth1}.

We conclude this section by emphasizing that balancing the requirement of strong  long range interactions with the necessity of minimizing neighboring site ground to Rydberg interactions fundamentally limits the number of qubits that can be directly connected without mechanical motion, swap gates, or conversion to flying qubits. Using a set of realistic parameters and a gate error target of $E=0.001$ the limit is about 500 qubits in a 2D geometry and 7500 qubits in a 3D geometry. While this number is not extremely large it may be sufficient for 
performing quantum simulations that are intractable on classical computers, and it  significantly exceeds the limit of present trapped ion based approaches for which $N_{\rm max}\stackrel{<}{\sim}10$ unless mechanical motion is invoked. 

\subsection{Maximum number of ensembles}

One of the primary challenges associated with building a 500 qubit device is the need to prepare 500 sites with a single atom in each. 
In the following we will discuss an approach to increasing  $N_{\rm max}$ based on collective encoding of 
an $N=60$ qubit  register at each of a smaller number of sites.  Collective encoding removes the requirements of individual qubit addressing and preparation of sites with single occupancy. The scaling laws are therefore different  than those of Eqs. (\ref{eq.Nmax2}). Let us assume 
that $K$ atoms are used for register encoding in each ensemble  and that the atoms are contained in a  spherical volume of diameter $D_K.$ We envision an architecture where gates can be performed within one ensemble and also between two ensembles in the array shown in Fig. \ref{fig.array}. For  inter-ensemble  gates we will access Rydberg states with $n=100$ and therefore the atom spacing inside each ensemble must respect the limit of $D_{\rm min}=0.7~\mu\rm m$ found above. 

As will be discussed in Sec. 
\ref{sec.collisions} we propose to base each ensemble on  $K\sim 100$ atoms trapped in a lattice with 
periodicity $D_{\rm min}$ and a filling factor $f=0.5.$  These numbers imply a sphere with diameter 
$D_K= \left(\frac{6}{\pi} \right)^{1/3}  D_{\rm min} (K/f)^{1/3} = 5.1~\mu\rm m.$ The relatively small $5.1 ~\mu\rm m$ maximum separation between atoms inside one ensemble allows intra-ensemble gates to be based on the blockade mode of operation which is insensitive to the precise value of the atom separation, while inter-ensemble gates can be based on the limit of $\Delta_{\rm vdW}\ll \Omega$ discussed in the preceding section. 
An array of ensembles, each with $D_K=5.1~\mu\rm m$ placed on a 2D grid with periodicity $D=5.3~\mu\rm m$ gives $k_{1, \rm eff}=10.$ and Eq. (\ref{eq.Nmax2}a) then predicts that $17$ such sites can be connected.
With each site containing $N=60$ qubits this forms the basis for a 1000 qubit scale processor. 
 We will discuss in detail in Sec. 
\ref{sec.collisions} an efficient  method by which loading of the ensembles can be achieved in parallel.

\begin{figure}
\centering
\includegraphics[width=6cm]{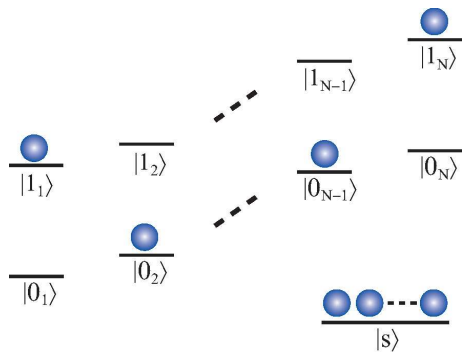}
\caption{(color online) Qubit encoding in the symmetric states of an ensemble of
$(2N+1)$-level systems. $|s\rangle$ denotes the reservoir state. The
figure depicts the state $|10\ldots 01\rangle$.}
\label{fig.levels}
\end{figure}

\section{Scaling by collective encoding}
\label{sec.holmium}

The limit on $N_{\rm max}$ found in the preceding section can be increased by a factor of $N$ using collective encoding of a $N$ qubit register in an ensemble of $K>N$ atoms at each lattice site.
We have recently described two different approaches to encoding a register in an ensemble of 
atoms each with $N'>N$ stable ground states\cite{ref.molmer1,ref.molmer2}. 
 In the approach shown in Fig. \ref{fig.levels} the $|0_i\rangle$ and $|1_i\rangle$ states of each register qubit $i$ are associated with single collective excitations of different hyperfine ground states. 
While this approach requires two internal states per qubit, which is not the most efficient  encoding 
possible\cite{ref.molmer1}, it has the distinct advantage that error correction protocols can be implemented in a fairly straightforward way\cite{ref.molmer2}.

The register size is limited by the number of stable internal states. 
In \cite{ref.molmer1} we discussed the use of  Cs which has 16 hyperfine ground states, and would allow for a 7 qubit register using the encoding of Fig. \ref{fig.levels}. Other atomic species have many more stable ground states. The rare earth atoms in particular with unfilled $f$ shells have large nuclear and electronic spins, and therefore many hyperfine ground states. The rare earths also have  large hyperfine splittings and ground state multiplet  splittings which turns out to be useful for qubit preparation and readout.   In this section we discuss the prospects for encoding a 60 qubit register in Ho which has 128 hyperfine ground states, more than any other stable atom. 
 
\begin{figure}
\centering
\includegraphics[width=8.7cm]{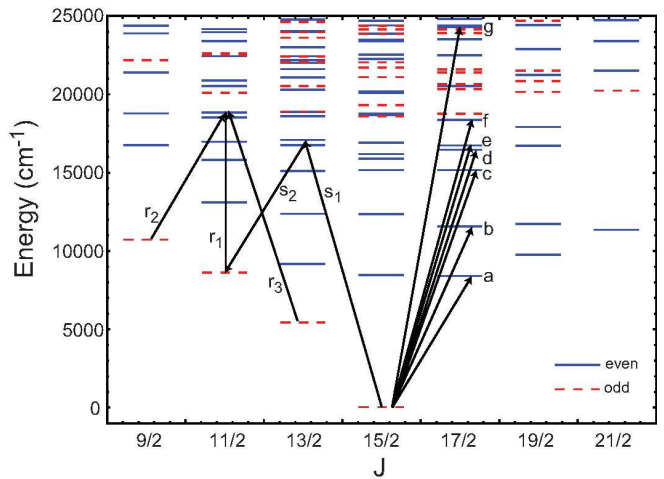}
\caption{(color online) The known\cite{ref.nist} 111 levels of Ho below $25,000~\rm cm^{-1}$ are shown with odd parity levels indicated by a dashed line. Transitions suitable for cooling and trapping (a-g), shelving to the $J=11/2$ metastable ground multiplet ($s_1, s_2$),   and  readout by resonance fluorescence  ($r_1, r_2, r_3$), are indicated.   }
\label{fig.Hocooling}
\end{figure}

\subsection{Register encoding in hyperfine ground states of Ho}

\begin{figure*}[!t]
\centering
\includegraphics[width=14.cm]{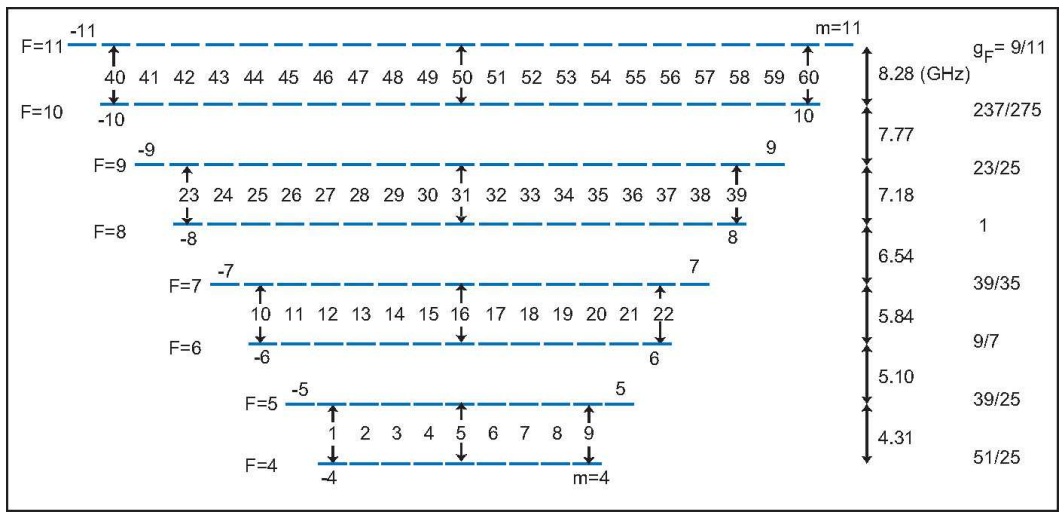}
\caption{(color online) Hyperfine structure of the Ho $4f^{11}6s^2 (^4I_{15/2})$ ground state. 
Qubit assignments 1-60 are indicated together with hyperfine splittings and $g$ factors. }
\label{fig.Ho}
\end{figure*}

To start let us recall some basic facts about the structure of the rare earth Ho shown in Fig. \ref{fig.Hocooling}.  
There is one stable isotope $^{165}$Ho which has nuclear spin $I=7/2$ and a ground electronic configuration 
$4f^{11}6s^2$. The ground state term is $^4$I$^{\rm o},\, J=15/2$, giving 128 hyperfine states with $4\le F\le 11.$
Transitions suitable for laser cooling and trapping are labeled a)-g). Ho is characterized by very large term dependent shifts and as seen in Fig. \ref{fig.Hocooling} there are three odd-parity metastable levels in the ground multiplet with $J=13/2,11/2,9/2$ at energies of $5420, 8605,$ and $10700~\rm cm^{-1}$. These auxiliary levels  will be used for qubit readout.

The ground and low-lying excited states of Ho are characterized by large hyperfine splittings  which are convenient for qubit encoding. 
The hyperfine constants of the ground state multiplets are known with high accuracy\cite{ref.Hohfs1,ref.Hohfs2} leading to the splittings shown in Fig. \ref{fig.Ho}.
A two-state per qubit encoding giving 60 hyperfine qubits is shown in Fig. \ref{fig.Ho}. 
Each bit is associated with the levels $|F,m\rangle\, (=|0_i\rangle)$ and $|F+1,m\rangle\, (=|1_i\rangle)$
with $F=4,6,8,$ or $10.$ We see that all states, except for the 8 with $F$ odd and $m_F=\pm F$ are assigned to qubits. The energy splittings between the 0 and 1 states of each qubit range from 
4.3 to 8.3 GHz. The qubit phase is sensitive to magnetic fields since the $g_F$ Land\'e  factors are different for all $F$ levels. The values indicated in the figure assume $L-S$ coupling and no configuration mixing, 
which is an accurate description for the ground state multiplet\cite{ref.Hohfs2}. A stable magnetic environment is therefore necessary to prevent qubit dephasing. One or more of the unused states can be assigned to the reservoir $|s\rangle.$ A convenient choice is to use $|s\rangle=|11,11\rangle$ 
this state can be populated by pumping on transition f) at $545.3~\rm  nm$ with $\sigma_+$ light.

\subsection{Cooling and trapping}

The rare earth atoms including Ho  have been magnetically trapped and collisionally cooled with He buffer gas to mK temperatures\cite{ref.Doyle}. This approach requires strong  magnetic 
fields of several T and is not directly suitable for qubit storage. There has also been recent progress in  laser cooling and magneto-optical trapping of the rare earth Er without the need for a repumping laser, despite the lack of a closed cycling transition\cite{ref.Ercooling}. It has been argued convincingly that
the large magnetic moments of Er contribute to the success of laser cooling and trapping, despite the fact that a cycling transition was not used, due to the possibility of magnetic trapping in a moderately strong quadrupole field. The spectrum of Ho is very similar to Er as concerns laser cooling possibilities, and if anything looks even more favorable due to the presence of sufficiently strong closed cycling transitions for cooling. 

The transitions labeled a)-f) in Fig. \ref{fig.Hocooling} are all closed cycling transitions 
between the $J=15/2$ ground state and $J=17/2$ excited states with vacuum wavelengths of $1193, 867.3, 660.9, 608.3, 598.5,$ and $545.3 ~\rm nm$. 
The only dipole allowed decay path from the upper level of these transitions is back to the ground state so we expect leakage out of the cycling transition to be negligible.
Transitions a)-d) have an upper level configuration of $4f^{10}5d6s^2$ which is dipole allowed, but has small oscillator strengths to the ground state. Transitions e) and f)  
have an upper level configuration of $4f^{10}6s6p$ and are coupled more strongly with the ground state. 
 The transition linewidths are however only known for d) and e) which have\cite{ref.gorshkov,ref.lawler,ref.nave} 
$(\gamma_d,\gamma_e) = (0.25,0.92)\times 10^6~\rm s^{-1}$ which give  Doppler cooling limits of $0.95, 3.5~\mu\rm K.$
These are attractively low temperatures but the lines may be too narrow to allow efficient capture from a background thermal vapor. Transition f) may have a larger linewidth, but this will have to be studied experimentally. If not there is the possibility of cooling on the 
very strong transition labeled g) to the level 
$4f^{11}(^4I_{15/2}^{\rm o})6s6p(^1P_1^{\rm o }), J=17/2$
at $410.5~\rm nm$  which appears analogous to the strong blue line used for cooling of Er to subDoppler temperatures\cite{ref.Ercooling2}. This transition has a linewidth of $\gamma = 204\times 10^6~\rm s^{-1}$ and a Doppler cooling limit of $780~\mu\rm K.$  The subDoppler temperatures observed in Er may have been due to a fortuitous  coincidence of ground and excited state $g$-factors, and may not occur in Ho. Nevertheless the prospects for laser cooling of Ho look very   promising either in a single step using lines d) or  e), or by using g) first to have a large capture range, followed by deep cooling on d). 

We note that the Er cooling experiments were successful without the use of a repumper, even for $^{167}$Er which has ground state hyperfine structure . The same may be true for Ho, although for quantum information applications it will be necessary to prepare the sample with all population in the reservoir state $|s\rangle.$ This can be done either by shelving the $|s\rangle$ level in one of the 
metastable ground multiplets, and then blowing away the unwanted atoms with unbalanced resonant light, or by adding repumper frequencies to deplete the lower hyperfine levels during cooling. With either approach it may be necessary to have as many as 7 repumping frequencies to depopulate all  hyperfine levels
$4\le F\le 10.$

After laser cooling and magneto-optical trapping we can transfer the atoms into tightly confining optical traps for qubit manipulation.  
 In order to hold ground and Rydberg state atoms in the same   trap it is preferable to use blue detuned trapping light
such that  both the ground state and the Rydberg electron have negative polarizabilities and can 
be simultaneously confined. 
A blue trap also minimizes spontaneous scattering of trapping light and is optimal for achieving long ground 
state coherence times\cite{ref.swth1}. Dipole allowed excitation from the Ho ground state to an even parity level is possible at a large number of wavelengths.  The use of a short wavelength transition allows for tighter spatial focusing and confinement.

\begin{figure}
\centering
\includegraphics[width=8.7cm]{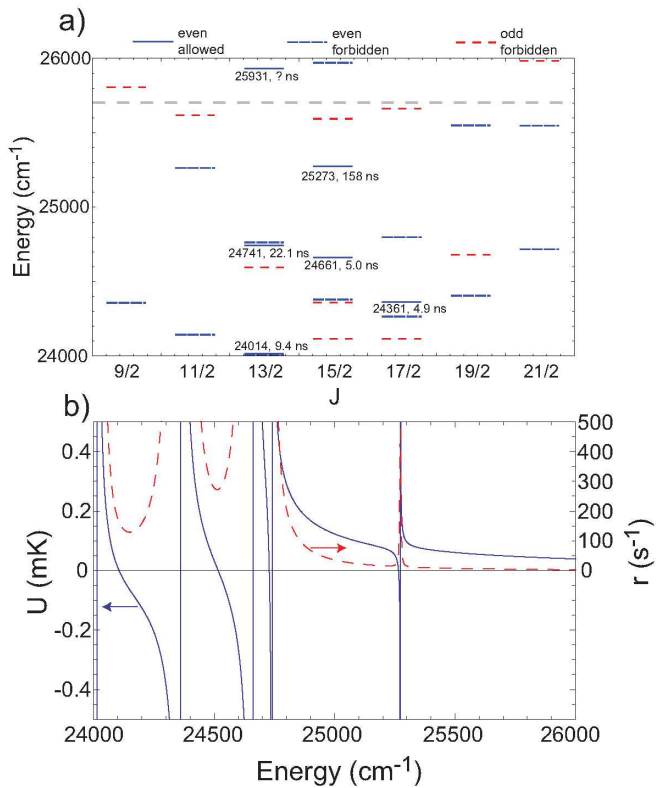}
\caption{(color online) Levels between 24000 and $2600~\rm cm^{-1}$ that are connected by dipole allowed transitions with the ground state are shown in solid blue, and dipole forbidden levels are dashed. 
The dipole allowed levels are labeled with their energy and lifetime.  Part b) shows trap depth (solid curve) and scattering rate (dashed curve) for light of the indicated energy.   }
\label{fig.Hofort}
\end{figure}

Figure \ref{fig.Hofort} shows all levels between 24000 and $2600~\rm cm^{-1}$, including the upper level of the strong g) transition discussed above at $24360.81~\rm cm^{-1}.$ Although this transition is a good candidate for cooling it turns out not to be directly convenient for a blue detuned optical trap due to the proximity of a second, almost equally strong transition to the $4f^{11}(^4I_{15/2}^{\rm o})6s6p(^1P_1^{\rm o }), J=15/2$ level at $24660.80~\rm cm^{-1}.$ 
 This line has very similar strength to g) with\cite{ref.nave} $\gamma=200\times 10^6 ~\rm s^{-1}.$  
In order to accurately estimate trapping conditions we have accounted for all the levels in Fig. \ref{fig.Hofort} which have dipole allowed transitions to the ground state.
 We can estimate the possible trap depth and photon scattering rate by assigning the indicated lifetimes to a single transition to the ground state for each level. This is not a bad approximation since there are no odd parity states between the ground multiplets and $18572~\rm cm^{-1}$, so the indicated lifetimes are roughly equal to the inverse of the decay rate $\gamma$ to the electronic ground state for each level. 
We also neglect the level with $J=13/2$  at $25931~\rm cm^{-1}$. The lifetime is unknown for this level,
presumably because it has configuration $4f^{10}5d6s^2$ and is only weakly coupled to the ground state. 
There are also additional dipole allowed transitions to levels at higher energy, the first one being  
$4f^{11}6s6p,\, J=15/2$ at $26958~\rm cm^{-1}.$ Again the lifetime is unknown, and presumably there is only weak coupling to the ground state. 
 
Using the two-level formulas for the optical potential $U=\frac{3\pi}{2}\frac{c^2}{\omega_a^3}\frac{\gamma}{\Delta}I,$ ($\omega_a$ is the transition frequency, $\Delta=\omega-\omega_a$ is the detuning, and $I$ is the intensity) and the scattering rate $r=\frac{3\pi}{2}\frac{c^2}{\hbar\omega_a^3}\frac{\gamma^2}{\Delta^2}I$, and summing the contributions from each of the allowed levels in Fig \ref{fig.Hofort}a) we obtain the 
optical potential and scattering rate curves shown in Fig. \ref{fig.Hofort}b) for 
 $5~\rm mW$ of power focused to a Gaussian waist  ($1/e^2$ intensity radius) of $w=5~\mu\rm m.$
  We see that the ratio of well depth to scattering rate is optimized at energies above $25500~\rm cm^{-1}.$
Tuning to $25700~\rm cm^{-1}$ ($389~\rm nm$) gives a trap depth of $100~\mu\rm K$ 
and a spontaneous scattering rate of just under $10~s^{-1}$. Note that the scattering rate will be further reduced in a blue detuned trap by approximately $k_B T_a/U$ where $k_B$ is the Boltzmann constant and $T_a$ is the atomic temperature.   We conclude that optical trapping of cold Ho looks relatively straightforward with low power light in the near uv part of the spectrum.

\subsection{Initialization, single qubit operations, and measurements}

Assuming that the task of cooling and trapping Ho has been completed  let us now consider how a qubit register can be initialized and manipulated. Referring to Fig. \ref{fig.Hocooling} we see that any of the transitions a)-f) is a closed  cycling transition. Using $\sigma^+$ polarized light and including repumping frequencies 
to depopulate all hyperfine levels with $F<11$ will result in initialization of  all atoms  into the reservoir state $|s\rangle=|11,11\rangle.$  The Ho  ground state has vanishing configurational mixing with other levels\cite{ref.Hohfs2} so the limiting factor as regards the efficacy of state preparation will be due to mixing of the upper state of the transition used for pumping, as well as Raman transitions into lower hyperfine states. As mentioned above, one approach to maximizing the state purity is to shelve the $|s\rangle$ state in an excited metastable level of the ground electronic configuration using transitions $s_1, s_2$ in Fig. \ref{fig.Hocooling}  at $586.2$ and $1183~\rm nm$ respectively, which proceed via the $4f^{11}(^4I_{15/2}^{\rm o})6s6p(^3P_1^{\rm o}), J=13/2$ excited state. We can then push away any unwanted population left behind in the other ground state levels using unbalanced force from light resonant with the e) transition.

Having prepared all atoms in $|s\rangle$  the qubit register must be initialized to a fiducial state. 
This can be done using sequences of stimulated Raman transitions starting with the state ``furthest" away from $|11,11\rangle$ and 
then working backwards. Since the register states are collective states the initial step must involve a Rydberg interaction, after which additional steps can be taken using stimulated Raman via 
for example transition e). Specifically, starting in $|s\rangle$  we prepare the collective state with unit excitation in 
$|0_{60}\rangle=|10,10\rangle$ via  Rydberg blockade. This state can then be mapped onto $|0_1\rangle=|4,-4\rangle$ 
using 7 steps of two-photon stimulated Raman via  the cycling transition e). 
We then prepare $|0_2\rangle, |0_3\rangle,...$ and so on.

Single qubit rotations  $|0_i\rangle\leftrightarrow |1_i\rangle$ on register bit $i$ are then straightforwardly performed using two-photon stimulated Raman beams tuned close to transition e). The qubit state  energy separations 
range from 4.3 GHz for bits 1-9 ($F=4\rightarrow 5$), to 8.3 GHz for bits 40-60 ($F=10\rightarrow 11$). In order to distinguish between different bits with the same value  of $F$ we rely on Zeeman shifts due to an 
external magnetic field. Since the $g_F$ factors are different for each $F$ level this is possible.

Qubit readout can be performed using shelving. We transfer the $|1_i\rangle$ part of bit $i$ to the metastable $J=11/2$ ground state multiplet using two-photon stimulated Raman with fields $s_1,s_2$ as described above, 
the only change being that the frequency $``s_1-s_2"$ must be tuned to be resonant with the
transition from  level $|1_i\rangle$ to a level in the $J=11/2$ multiplet.
The $J=11/2$ level can then be coupled to $4f^{11}(^4I_{15/2}^{\rm o})6s6p(^3P_2^{\rm o })$ using the $r_1$ transition at  $878.8~\rm  nm$ to generate fluorescence photons. The upper level only has dipole allowed decay paths to the 
metastable $J=9/2,11/2,13/2$ levels shown in Fig. \ref{fig.Hocooling}, all of which can be pumped back up to the same upper level. This requires two additional repumpers $r_2, r_3$ at 1231 and $746.2~\rm  nm$. If the measurement gives a null result the register bit  is projected 
into $| 0_i\rangle$ and we are done. 

If the measurement result is $| 1_i\rangle$ then the register bit will have to be restored  for further processing. The atom can be optically pumped into the $|J=11/2, F=9,m_F=9\rangle$ state using $\sigma^+$ light on transition $r_1$ and then  coherently transferred back to $|s\rangle$ using transitions $s_1$ and $s_2$. 
 All bits  $j>i$ can then be swapped down to fill the hole at bit $i$ 
($|0_{i+1}\rangle \rightarrow |0_i\rangle,$
$|1_{i+1}\rangle \rightarrow |1_i\rangle,$
$|0_{i+2}\rangle \rightarrow |0_{i+1}\rangle,$
$|1_{i+2}\rangle \rightarrow |1_{i+1}\rangle,$
 etc.), and the mapping between information and register bits can be relabeled to account for the change. 
Bit 60 can then be reinitialized from $|s\rangle$ via a blockade operation and the computation can continue. 
Alternatively, all the swap operations can be avoided by restoring the atom to the reservoir $|s\rangle$ and then performing the first blockade mediated operation from $|s\rangle$ back to a state in $J=11/2$. From there the state can be  coherently transferred
to  $|F',m_{F'}\rangle$ in the $J=11/2$ level using stimulated Raman. The measured register bit $|J=15/2,F,m_F\rangle$ can thus be restored after the measurement to $|J=11/2,F',m_{F'}\rangle$ where  $|F'-F|\le 2$ and $|m_{F'}-m_F|\le 2.$  
This state can then be returned to restore the register bit $|J=15/2,F,m_F\rangle$ using stimulated Raman on $s_1, s_2.$ This eliminates the need for swapping many register bits at the expense of an additional laser to couple $J=11/2$ to Rydberg levels. 
Using either approach we thus have a protocol which 
allows for resetting of a measured qubit so that a computation can proceed as long as there is no physical loss of atoms.

Finally, there are two additional issues which should be mentioned in connection with single qubit operations and measurements.
The first is that the optical trapping discussed above using light at $389~\rm nm$ will not produce the 
same trapping potential for the $J=9/2,11/2,13/2$ states. This can potentially be solved by adding additional  beams that are each tuned to create an attractive potential for these states, but are far off-resonant with respect to the ground state, and do not disturb the unmeasured part of the register. Since these beams are only needed during a brief measurement some amount of photon scattering can be tolerated in $J=9/2,11/2,13/2$, and the choice of wavelengths will not be as constrained as for the calculations shown 
in Fig. \ref{fig.Hofort}. It is also feasible to consider readout based on cross entanglement between 
atoms of different species\cite{ref.RbCs}, which removes the requirement of using the metastable levels entirely. A detailed study of this possibility is outside the scope of this work.

The second issue is related to Zeeman selectivity. The ground state hyperfine separations between levels with neighboring  values of $F$ range from 4.31 to 8.28 GHz, so selective shelving  to or from the auxiliary 
 levels as well as Rydberg levels are  well resolved as regards different $F$ values. For each value of $F$, bits with different $m$ can be isolated by applying a magnetic field $B$ to give shifts $\Delta U_{|F,m\rangle}=g_F \mu_B B m.$ The $g_F$ values range from 0.82 for $F=11$ to $2.04$ for $F=4.$ Moderate fields of  under
2.2 Gauss will thus be sufficient to get 2.5 MHz of separation between all qubits as regards shelving and Rydberg excitation. This allows the use of greater than $100~\rm kHz$ Rabi frequencies while keeping the probability of  population transfer of a nonaddressed bit below the $10^{-3}$ level in the worst case, and much less for the majority of register bits. 

There remains, however, the problem of selectively addressing bits for single qubit rotations. In this case the resonance condition scales as $\Delta U_{|F,m\rangle}-\Delta U_{|F-1,m\rangle}=(g_F-g_{F-1}) \mu_B B m.$
The worst case is for bits  40-60 since $g_{10}=0.86$ and $g_{11}=0.82.$ The selectivity between bits with neighboring $m$ values is thus only $0.044 \mu_B B$ or about $61~\rm kHz/G.$ To run single qubit operations at a $100~\rm  kHz$ Rabi frequency and have errors at the $10^{-3}$ level on  a neighboring bit 
requires a detuning of about $3 ~\rm MHz$ or  about a $50~\rm G$ field. While this is not particularly large we simultaneously require small dephasing 
on the most sensitive bits 1 and 9 which see  a  big differential shift of $135~\rm MHz.$ Any known differential rotation can be accounted for, so the feasibility of Zeeman selectivity without inducing unwanted dephasing rests on the ability to create a very low noise bias magnetic field. Such a large field only needs to be turned on for a time given by the inverse Rabi frequency or about 
$10~\mu\rm s$. To keep the dephasing error at $10^{-3}$ of a radian would require a field stability of 
$\sim10^{-3}/1350\sim 10^{-6}.$ This is not impossible but will be a technically challenging requirement. 

There are several possible approaches to mitigating this stability requirement. 
Since shelving is well resolved by much smaller magnetic fields than those needed to resolve single qubit rotations, we can shelve bits 
$|0_i\rangle, |1_i\rangle$ to the  metastable $J=11/2$ level, do the rotation there, and then return them to the ground state levels. 
Alternatively, to perform a rotation on a qubit encoded in $F=10,11$ states we first swap the bit with a bit encoded in $F=8,9$ ($11\rightarrow 9$ and $10\rightarrow8$), relabel the information, and then perform the rotation on the bit in $F=8,9.$ 
This has the advantage that the swap operation has a larger selectivity governed by the $g_F$ of the lower  $F$ states.  Swapping in this way from $F=10,11$ to $F=8,9$ reduces the bias field requirement by a factor of $(g_9-g_{11})/(g_{10}-g_{11})=2.3.$

\subsection{Rydberg interactions}

Finally we need to consider Rydberg interactions in Ho. The asymptotic scaling of Eq. (\ref{eq.C6p}) is expected to hold  at large $n$ for any atomic species with singly excited Rydberg states so the estimates 
found in Sec. \ref{sec.lattice} should  remain valid, although the precise numerical values will require adjustment.  We know of no principal reasons why Rydberg blockade analogous to what has been studied in the alkalis\cite{ref.dipoleblockade} should not be possible but 
a detailed characterization  remains a topic for future investigation.
We are aware of only one
experimental  study  of the Rydberg structure of Ho\cite{ref.Horydberg} where  the Rydberg series corresponding to excitation of  $4f^{11}6snp$ was observed by collisional ionization. Resolved Rydberg levels up to $n\simeq 47$ were seen.  
As these are even parity states excitation from the ground state requires either one $\sim 210~\rm nm$ 
photon or a three step excitation at longer wavelengths. The $4f^{11}6sns$ or $4f^{11}6snd$ series which are accessible by two step excitation are a second possibility.

The largest uncertainty as regards the feasibility of Rydberg gates in Ho concerns the Rydberg state lifetimes. In Sec. 
\ref{sec.lattice} we assumed $\tau\sim n^2$ blackbody scaling. If the Ho Rydberg series are perturbed
by interactions between the valence  and core electrons  the lifetime could be substantially altered. 
If strong core - valence interactions do occur they are likely to be specific to particular $n$ values, so that with judicious choice of the Rydberg level it should be possible to minimize the impact of series perturbations.


\section{Preparation of a lattice of collision suppressed ensembles}
\label{sec.collisions}

In light of the principle feasibility of collective encoding in Ho discussed above, it is interesting to examine how an array of ensembles can be efficiently prepared in the geometry shown in  Fig. \ref{fig.array}. 

\begin{figure*}[!t]
\vspace{-.1cm}
 \centering
  \begin{minipage}[c]{15.5cm}
\includegraphics[width=15.5cm]{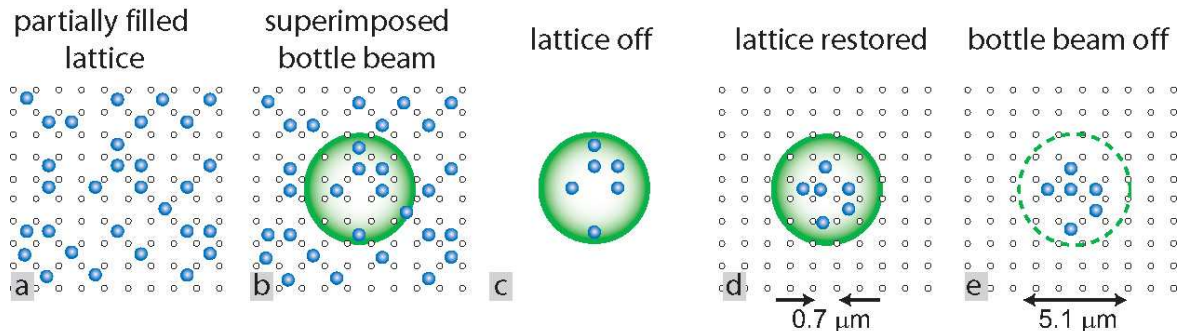}
 \vspace{-.9cm}
 \caption{Protocol for loading small ensembles into a patterned lattice region. The small open circles are the repulsive lattice sites, and the large green circle is the outline of the bottle-beam trap. 
The lattice is not drawn to scale for clarity. See text for details. 
}
\label{fig.loading}
\end{minipage}
\end{figure*}

The use of a many atom ensemble at each site, instead of requiring deterministic loading of single atoms,
immediately removes one of the prime challenges of neutral atom quantum computing, which has been  
the difficulty of preparing a singly  occupied lattice of optically resolvable sites.  There has been a considerable amount of work on this problem based on  several different solutions. An elegant approach  
relies on the BEC to Mott insulator transition\cite{ref.becmott}  as has been demonstrated
in recent experiments\cite{ref.blochbecmott}.  Unfortunately the transition only works in finite time with a very short period lattice which does not give optically resolvable lattice sites. 
This has been partially addressed by the  demonstrated transfer of the insulating state to a longer period lattice\cite{ref.portoloading}.
While the Mott insulator transition at
finite temperature does not result in a lattice that is filled with perfect fidelity there are also possible solutions
to purifying the lattice\cite{ref.weissfix}. These purifying steps can also potentially be used on a lattice which is imperfectly filled directly from a magneto-optical trap (MOT), without needing a BEC phase\cite{ref.weissload}. One of us has also proposed a deterministic  loading scheme\cite{ref.sw1} which relies on Rydberg blockade to remove all but one atom from a multiply filled site. This can potentially be implemented in parallel on a large number of sites.

The ensemble based approach described here is potentially much simpler since it does not require preparing sites with single atom occupancy. Based on the results of\cite{ref.molmer1} a 60 bit register, including error correction, needs an ensemble of $K>60$ atoms. For the purposes of the present discussion we will target a value somewhat larger say, $K\sim 100.$ These $K$ atoms should all be confined to a small volume to allow an effective Rydberg blockade and to maximize the total number of processor qubits. 
The disadvantage of a multi-atom ensemble is that high density samples suitable for Rydberg blockade experiments are susceptible to collisions which will drastically reduce coherence times. To solve this problem we propose to load the ensemble in a short period lattice, thereby eliminating collisional decoherence. 

The loading protocol is shown in Fig. \ref{fig.loading}. 
We start with Ho in a MOT, and assume Doppler cooling to a few $\mu\rm K$ using transition e). 
 The atoms are then  transferred into a 3D blue detuned lattice (3  pairs of beams with orthogonal polarizations, and frequency shifts to avoid unwanted interference) created with $\lambda=389 ~\rm nm$ light, as described above. The angle between each pair of beams is adjusted to give   a lattice spacing of $\Lambda= 0.7~\mu \rm  m$ which corresponds to a lattice site density of $n_{\rm lattice}= 2.9\times 10^{12}~\rm cm^{-3}.$ 
Magneto-optical trapping of Er has resulted in peak densities of $n_a\simeq 2\times 10^9 ~\rm cm^{-3}$
and we expect similar baseline results for Ho. 
It is unknown what densities can be achieved using transient magnetic compression and/or evaporative cooling, but by analogy with experience from the alkalis where evaporation has produced small 
samples with densities above $10^{15}~\rm cm^{-3}$\cite{ref.walkerevap} preparing 
a few thousand atoms with $n_a\simeq 1.5\times  10^{12}~\rm cm^{-3}$  appears realistic. 
Loading such a sample into the above lattice would give a filling fraction of $n_a/n_{\rm lattice}\sim 0.5.$ Assuming Poissonian loading statistics the probability of double site occupancy at this filling fraction is $0.09$ which implies that for  $K\sim 100$ about 18 sites will have more than one atom. The atoms in these sites  will rapidly redistribute or be lost from the trap due to hyperfine changing collisions leaving a sample with slightly smaller $K$ but only single atom occupancy at all sites.

We then superimpose a blue-detuned bottle beam trap\cite{ref.larry1}  and drop the lattice to remove atoms outside the region defined by the bottle beam (frames b,c in Fig. \ref{fig.loading}).  The lattice is then restored and the bottle beam turned off (frames d,e) which leaves us with $K$ atoms in the lattice, in a well defined spatial region. We can achieve an average value of $\bar K = 100$ using 
$n_{\rm a}\sim 1.5\times 10^{12}~ \rm cm^3$ and a spherical bottle beam trapping region with diameter $D=5.1 ~\mu\rm m$ (the bottle beam trapping potential is not far from spherical for such small trap volumes). 
The probability of doubly occupied sites, and hence collisions, will be somewhat higher than in the initial lattice state since in the bottle beam transfer phase when the lattice is dropped the atoms will tend to fall towards the center and compress.  
This effect will be smaller than in a harmonic trap since our bottle beam design provides a quartic transverse potential, $U\sim r^4$, which provides only weak radial compression. The potential is harmonic axially, but  by judicious choice of the time window for the bottle beam transfer $\Delta t\sim p t_{\rm vib},$ where $t_{\rm vib}$ is the axial vibrational period, and $p$ is an integer multiplier, compression effects should be minimized.

Following the above procedure we can prepare a localized region with $\bar K\sim 100$ atoms. This can readily be done in parallel at multiple sites by superimposing an  array of bottle beams on the 3D lattice. Poissonian loading statistics will imply $\sim10\%$ variations in $K$ from site to site, and hence  variations in Rabi frequencies for qubit operations involving Rydberg levels.
For the first encoded qubit which is supported by $K\sim 100$ atoms the Rabi frequency variations will be 
$\sqrt{1.1}\sim 1.05$ or $\pm 5\%$.  
For the last encoded qubit which is supported by $K-59\sim 40$ atoms the Rabi frequency variations will be 
$\sqrt{1.25}\sim 1.12$ or $\pm 12\%$.  
 These variation will not prevent high-accuracy gate operations since composite pulse sequences can be readily used to remove the dependency on Rabi frequency variations. 
We have thus arrived at the situation shown in Fig. \ref{fig.array} with an array of closely spaced ensembles. If the diameter of each ensemble is about $5.1~\mu\rm m$ then we might space them in a plane by $D\sim 5.3~\mu\rm m$ using addressing beams with flatter than Gaussian profiles to minimize site to site cross talk. The scaling arguments of  Sec. \ref{sec.lattice}  imply we can connect 
roughly $17$ ensembles. The total processor size is thus $60\times 17=1020$ directly coupled qubits.  

\section{Discussion}
\label{sec.discussion}

We have presented a path towards a neutral atom gate array that could allow as many as a thousand qubits 
to be interconnected without resorting to mechanical motion, swap chains, or interconversion between stationary and flying qubits. The scaling relies on optimization of qubit array and Rydberg parameters to allow 
up to 500  sites to be directly coupled in a planar geometry. The estimate was based on realistic parameters, including gate errors of $E=0.001$ and a maximum principal quantum number of $n=100$. This limit may be too conservative, and the ultimate limit in a planar geometry may be higher than predicted here. 
Nevertheless manipulating single atoms in such a  large number of sites presents a difficult experimental challenge. 
We can avoid the overhead associated with individual addressing and single atom loading at each site  by invoking collective encoding. 
Using Ho atoms this gives a  multiplying factor of 60 per site. Since the ensemble sites are larger than those holding a single atom we arrive at a limit of about 17 sites, each with 60 collectively addressed qubits, giving a 1000 qubit scale device. Our estimates are based on available spectroscopic data for Ho. More information, particularly concerning the Rydberg states of Ho, will be needed to further refine the limits discussed here. 

Another attractive application of collective encoding would be to use a single ensemble with a 60 qubit register as a quantum repeater, or as part of a hybrid quantum computation scheme involving small, optically connected registers\cite{ref.scalablerepeater}.  The 60 qubits are sufficient for 8 logical qubits, each built from 7 physical qubits, or, for example, 5 logical qubits, each built from 5 decoherence free subspace encoded qubits, that  in turn require 2 physical qubits each\cite{ref.rydbergdfs}. In this way 50 physical qubits 
would enable 5 error corrected and potentially low decoherence logical  qubits in one ensemble.
The  ensemble qubits can be efficiently mapped to and from  
photonic bits using blockade mediated preparation of excited states\cite{ref.sw1}, and the large choice of transitions in Ho provides flexible opportunities for coupling to wavelengths compatible with optical fiber transmission.

There are undeniably many challenges associated with collective encoding, particularly in rare earth atoms that are relatively poorly studied and have not been widely used for laser cooling. A large number of lasers of different wavelengths and frequencies are required for  the various internal state manipulations.
In some sense we have transferred the complexity of moving the information about spatially, as in current approaches to scalable ion traps, to the problem of dexterously moving information between internal states. 
Although  the overall complexity required to build a scalable quantum processor will remain very high, 
we believe it is   worthwhile to explore a wide range of approaches.  
 Indeed,  we are not aware of any  approaches to building a thousand qubit scale quantum logic device that are simple.  

This work was supported by the NSF and ARO-IARPA.
M. S. thanks  Jim Lawler for helpful discussions on Ho spectroscopy.

\end{document}